\documentstyle[preprint,aps]{revtex}

\def\prd{Phys. Rev. D }
\def\prn{Phys. Rev.  }
\def\npb{Nucl. Phys. B }
\def\prl{Phys. Rev. Lett. }
\def\plb{Phys. Lett. B }

\def\ibd {{\it ibid. }}
\preprint{YUMS-97-26/SNUTP-97-129}
\begin{document}
\title{  The O(N) Nonlinear Sigma Model in the Functional Schr\"{o}dinger Picture}
\author{Dae Kwan Kim\footnote{e-mail:dkkim@phya.yonsei.ac.kr}
        and Chul Koo Kim\footnote{e-mail:ckkim@phya.yonsei.ac.kr}}
\address{Department of Physics and Institute for
         Mathematical Sciences, Yonsei University,
         Seoul 120-749, Korea}
\maketitle
\vspace{3cm}
\begin{abstract}
We present  a functional   Schr\"{o}dinger picture formalism   of the (1+1)-dimensional  
 
$O(N)  $  
 nonlinear 
sigma model.
The energy density has been  calculated to two-loop order using the wave 
functional of a  gaussian form, and  from which  the nonperturbative mass  gap of  the 
boson 
fields has been obtained. 
The functional Schr\"{o}dinger picture approach combined with the variational 
technique is shown
to describe the characteristics of the ground state of the nonlinear 
sigma model in a transparent way.
\end{abstract}
\draft
\vspace{0.5cm}
\newpage

\section{Introduction}

The ground state of the interacting quantum fields, in general, has a
complicated structure to investigate, rendering ordinary perturbation theories without 
much success. 
Especially, concerning spontaneous symmetry breaking and bound states, 
perturbative ground states mislead to wrong results \cite{nop}. 
In this   respect,  the   functional  Schr\"{o}dinger  picture   (FSP)  approach  with   
variational 
approximation is expected to 
be a useful tool to examine the nonperturbative aspects of
quantum field theory. 

In contrast to the usual perturbative expansion, the 
Schr\"{o}dinger picture approach has the merit that one does not need to specify 
a particular Fock basis for the  ground state of the Hamiltonian under 
consideration. Therefore, when there is no well defined Fock vacuum, this 
method appears to be a convenient choice \cite{vac}.

In particular, the nonlinear sigma (NLS) model \cite{sig,bar,jus} has a
nontrivial vacuum structure that is composed of particle-antiparticle pairs. 
The ground state is not easily tractable applying the usual perturbation 
expansion. Therefore,  it appears  as  a natural  candidate for   application of the  FSP 
approach.
The aim of this paper is to analyze the NLS model in the framework of the 
FSP method combined with a variational approach 
\cite{sch,res,pis,noh}, which is known to  go  beyond the perturbative  scheme in  some 
cases.. 
We will show 
that the nonperturbative phenomena like
the mass gap and asymptotic freedom can be described in the Schr\"{o}dinger picture in
a direct way.

The NLS model in lower dimensions has attracted much attention, 
since it has relevance to the low energy limit of QCD as well as condensed 
matter systems such as antiferromagnets.
The NLS model in two dimensions is classically scale invariant and 
asymptotically free \cite{sig,bar,jus}. According to Mermin-Wagner and Coleman
\cite{mem}, the continuous symmetry can not be broken in $1+1$ dimensions; the 
massless Goldstone bosons tend to acquire their masses. 
And it was shown that there is a mapping between the NLS model and the 
effective long wavelength action of the quantum Heisenberg antiferromagnet  
\cite{hei}. 

In section II, we briefly introduce the NLS model and its formulation in the FSP 
approach.
In the section III, we  first calculate the energy density to two-loop 
order using a gaussian-type wave functional, and then derive the mass gap 
for the boson fields by minimizing the energy density. We will show that in the 
NLS  model, the massive ground state is more stable than the massless one.  
In the last section, a brief summary and discussion of our results will be given. 
\section{ The Nonlinear Sigma Model in the Functional Schr\"{o}dinger Picture }
We start with the $O(N) $-invariant Lagrangian density \cite{sig,jus}
\begin{equation}
{\cal L}= \frac{{ 1}}{2 \lambda  } { \partial_{\mu} \Phi_{a}  \partial^{\mu} \Phi_{a} },
\end{equation}
where $N $ scalar fields $ \Phi_{a,} a=1, \cdots, N $, obey the constraint,
\begin{equation}
 \sum_{ {a }=1}^{ N} \Phi_{a} \Phi_{a} =1.
\end{equation}
This constraint makes the theory quite complicated, since the $N$  components
of the scalar field $ \Phi $ are mutually dependent on each other.
The coupling constant $ \lambda  $ is a measure of the 
strength of the self  interaction  of the  $N$  scalar fields $  \Phi_{a} $, and  a small 
value 
of $ \lambda  $ corresponds to a weak interaction.
The constraint  in Eq.(2)   means that  one degree   of freedom   among the   $N$   
variables, $ 
\Phi_{a}$, 
is not a real  dynamical variable \cite{con}.  Thus,  we follow the standard prescription 
to get 
rid of the {\it N}th
field $\Phi_{N} $ through the following nonlinear transformation \cite{bar}, 
\begin{equation}
 \Phi_{a} = \frac{{ \phi_{a}}}{( 1+  \phi^{2} /4  ) } ,
 \Phi_{N} = \frac{{( 1-   \phi^{2} /4  )}}{ (1+   \phi^{2} /4 ) } ,
\end{equation}
where $ \phi^{2} \equiv  \sum_{ { a}=1}^{N-1} {\phi_{a} }^2 $.
Substituting these expressions into the Lagrangian, we find the equivalent
Lagrangian involving the $N-1$ fields $\phi_{a} $,
\begin{equation}
{\cal L}=   \frac{{  1}}{2  \lambda   } \frac{{   \partial_{\mu} \phi_{a} 
\partial^{\mu} \phi_{a}  }}{(1+  \phi^{2} /4 )^{2} } .
\end{equation}
Here and in the following, the sum over the repeated indices are implied; 
otherwise, a comment will be given explicitly.

The NLS  model in terms of $\phi_{a} $'s has no mass 
parameter, so these fields are classically
massless Goldstone bosons. However, the Goldstone bosons originate from 
the breakdown of   the continuous    $O(N) $  symmetry,  which   can not occur in 
this case 
\cite{mem}. In the below, it will be seen that the 
Goldstone bosons become massive through
the quantum mechanical self-interaction. 

The canonical quantization procedure of the classical system require the conjugate
momentum of the field
$\phi_{a} $
, which becomes
\begin{eqnarray}
\pi_{a}  &=& \frac{{\partial  {\cal L}}}{ \partial {\dot{\phi}}_a } \nonumber \\ 
        &=& \frac{{1}}{\lambda } \frac{{\dot{\phi}}_a}{(1+\phi^{2} /4)^{2} } .
\end{eqnarray}
Thus, the Hamiltonian  is written in  terms of the  canonical variables,  $\pi_{a} $ and 
$\phi_{a} $ as
\begin{eqnarray}
  H & =& \pi_{a} \dot{\phi}_{a} -{ \cal L} ( \phi_{a} , \dot{\phi}_{a} ) \nonumber \\
       & =& \frac{{ 1 }}{ 2 } \lambda (1+\frac{{\phi^{2}}}{4}   )^{2} \: \pi^{2} + 
  \frac{{1}}{2 \lambda}
 \frac{{{ \phi^{\prime}}^{2}}}{ (1+\phi^{2}  /4   )^{2}} ,
\end{eqnarray}
where we used the conventions   $ \phi^{2} \equiv   \sum_{  {  a}=1}^{N-1} {\phi_{a}  
}^2 ,
\pi^{2} \equiv  
\sum_{ {   a}=1}^{N-1} {\pi_{a}  }^2 $   and ${   \phi^{\prime}}^{2} \equiv   \sum_{ { 
a}=1}^{N-1} 
(  d \phi_{a} / dx )^{2}  $.

In the quantum theory, the dynamical field variable 
$\phi_{a} (x)$ and $ \pi_{a} (x) $ become
field operators.
These $\pi_{a} $'s satisfy the canonical equal-time commutation 
relations with 
the $\phi_{a} $'s such that
\begin{equation}
[\phi_{a} (x), \pi_{b} (y)  ]_{x_{0} = y_{o} }= i \delta_{ab} \delta (x-y) .
\end{equation}
In the FSP representation of its quantum theory, the scalar field  
operator $\phi_{a} (x)  $ and its  conjugate momentum $\pi_{a}  (x) $ are realised  as 
\cite{sch,res,pis}
\begin{eqnarray}
\phi_{a} (x)  & \rightarrow &  \phi_{a} (x), \nonumber \\
\pi_{a} (x) & \rightarrow & - i \hbar \frac{{ \delta }}{ \delta \phi_{a} (x)} .
\end{eqnarray}
Now, we employ a Gaussian-type wave functional \cite{pis} with two 
variational function parameters $G_{ab}(x,y)  $ and ${\hat{\phi}}_{a} (x) $,
\begin{equation}
 \Psi [\phi] = \frac{{ 1}}{[{\rm \det}(2\pi \hbar G)]^{1/4}} {\rm \exp} \left [ 
- \int_{ x,y} ( \phi_{a} (x)- {\hat{\phi}}_a (x) ) \frac{{ G_{ab}^{-1} (x,y) }}{ 4 \hbar } 
 (\phi_{b} (y)- {\hat{\phi}}_b (y)  ) \right ] ,
\end{equation}
where repeated indices over a or b mean sums; otherwise, an explicit comment will  be 
given 
in the below.  
This wave functional may be seen to obey that
\begin{eqnarray}
\langle \Psi | \Psi\rangle &=&1,\nonumber \\
\langle \Psi | \phi_{a} (x)| \Psi\rangle &=&{\hat{\phi}}_a (x),\nonumber \\
\langle \Psi | \pi_{a} (x)| \Psi\rangle &=&0.
\end{eqnarray}
Here, ${\hat{\phi}}_{a} (x)   $  is the  expectation  value   of  the field    
operator $\phi_{a} (x) $, 
and the expectation value of the momentum operator has been chosen to be zero.
From the above trial wave functional, we readily obtain the following 
results, which are needed for calculation of the expectation 
value of the Hamiltonian,
\begin{eqnarray}
\langle\phi_{a} (x) \phi_{b}  (y)\rangle  &=& {\hat{\phi}}_a  (x)  {\hat{\phi}}_b  (y)  +  
\hbar  
G_{ab} (x,y) , \nonumber \\
\langle\pi_{a} (x) \pi_{b} (y)\rangle &=& \frac{{\hbar }}{4} G_{ab} ^{-1} (x,y) .
\end{eqnarray}  

\section{Large $N$ Calculations}

In order to obtain the expectation value of the Hamiltonian
with the gaussian trial wave functional 
$\langle \Psi |  H |  \Psi \rangle   $, it  is necessary  to handle the  field operator  in  
$( 
1+\phi^{2}  /4 )^{-2}  $ properly. For that purpose, we  expand it in terms of $\phi^{2} 
/4  $, 
which involves an infinite number of terms of even powers of $\phi_{a} (x) $ fields.

Thus, we have to evaluate terms as follows,
\begin{equation}
\langle \Psi | \overbrace{ \phi^{2} (x) \cdots  \phi^{2} (x)} \pi^{2} (x)| \Psi\rangle
 \: {\rm and }\:
 \langle \Psi   | \overbrace{  \phi^{2}  (x) \cdots    \phi^{2} (x)  }{  \phi^{\prime}}^{2}  
(x)| 
\Psi\rangle , 
\end{equation}
which can not be calculated in closed forms, except for 
some special limiting cases.
Due to  this inability,  we require   an approximation   scheme. Here, we  will  resort  
to the 
large-$N$ calculations \cite{col}.
The following example will show why   the large-$N$ approximation is useful  in  the 
present 
problem;
\begin{eqnarray}
\langle \Psi   |  \phi^{2}   (x)  \phi^{2}  (x)   \pi^{2}   (x)| \Psi\rangle &   =&   \left 
[     
({\hat{\phi}}_a 
{\hat{\phi}}_a + \hbar G_{aa}  )
( {\hat{\phi}}_b {\hat{\phi}}_b + \hbar G_{bb}  ) \frac{{ \hbar }}{ 4} G_{cc}^{-1}  \right 
] \nonumber \\
& &+ \hbar^{2} \left [   {\hat{\phi}}_a G_{ab} {\hat{\phi}}_b G_{cc}^{-1} - \frac{{ 3 }}{ 
4}{\hat{\phi}}_a  {\hat{\phi}}_a   \delta_{bb}   +   \frac{{ \hbar   }}{2}   G_{ab}G_{ba} 
G_{cc}^{-1} - \hbar  G_{aa} \delta_{bb} \right ] \nonumber \\
& & - 2 \hbar^{2} \left [  {\hat{\phi}}_a {\hat{\phi}}_a +  \hbar  G_{aa} \right ] ,
  \end{eqnarray}
where the three  terms in  the brackets on  the right-hand  side are  of  the order  of  
$N^{3}, 
N^{2}$, and $ N$ respectively.
Thus in the large $N$ limit, only the leading terms of $N^{3} $ order dominate.
 \begin{eqnarray}
\langle \Psi |  \phi^{2} (x)   \phi^{2} (x)  \pi^{2} (x)|   \Psi\rangle &  \stackrel {{\rm   
Large}~ 
N }{ 
\longrightarrow}  & ({\hat{\phi}}_a {\hat{\phi}}_a + \hbar G_{aa}  )
( {\hat{\phi}}_b  {\hat{\phi}}_b +   \hbar G_{bb}  )   \frac{{ \hbar }}{   4} G_{cc}^{-1} 
,
\nonumber \\
                                           & =&
\langle \Psi | \phi^{2} (x)| \Psi \rangle \langle \Psi  | \phi^{2} (x) | \Psi \rangle  \langle 
\Psi 
|\pi^{2} (x)| \Psi\rangle.
\end{eqnarray} 
Therefore it is clear that in  the large $N $ limit, the  expectation values of 
the composite operators become \cite{bar}
\begin{equation}
\langle  \Psi  |   \phi^{2} (x)   \cdots   \phi^{2}   (x) \pi^{2}   (x)  (  {\rm   or}
\;{\phi^{\prime}}^{2} (x))
| \Psi \rangle =
\langle \Psi  | \phi^{2}  (x)| \Psi   \rangle \cdots  \langle \Psi   | \phi^{2}  (x)  | \Psi  
\rangle  
\langle \Psi |\pi^{2} (x) ( 
{\rm or}\; { \phi^{\prime}}^{2} (x) )| \Psi\rangle.
\end{equation}
To construct  an $1/N   $ expansion  \cite{col} in  a systematic  way,   we  define a  
new  
parameter $g $  such 
that 
\begin{equation}
g \equiv \lambda N,
\end{equation}
where $g $ is fixed to be finite.
Thus, we are allowed to write the Hamiltonian expectation value in the following form
\begin{equation}
\langle \Psi | H |  \Psi \rangle = \frac{{ 1}}{  2} \frac{{ g}}{  N} \left[  1+\frac{{\langle 
\phi^{2} 
\rangle }}{4}   \right]^{2} 
\: \langle\pi^{2}  \rangle + \frac{{ 1}}{ 2} \frac{{ N}}{g }
    \left[  1+\frac{{\langle\phi^{2}    \rangle   }}{4}   \right]^{-2}     \langle  {     
\phi^{\prime}}^{2} 
\rangle ,
\end{equation}
where on the right side, the expectation has been taken with respect to 
the gaussian wave functional in Eq.(9). 
Using the results for $\langle \phi^{2} \rangle , \langle \pi^{2} \rangle $ in Eq.(11), 
this equation can be rewritten as 
\begin{eqnarray}
\langle   H   \rangle  &=&  \frac{{   1}}{ 2}    \frac{{ g}}{     N} [  1+\frac{{1}}{4} 
({\hat{\phi}}_a 
{\hat{\phi}}_a + \hbar G_{aa}  )]^{2} \frac{ \hbar }{4 } G_{cc}^{-1} \nonumber \\
            & &+  \frac{{ 1}}{  2}   \frac{{ N}}{g  }  [1+\frac{{1}}{4}  ({\hat{\phi}}_a 
{\hat{\phi}}_a + \hbar G_{aa}  )   ]^{-2}
 ( \nabla  {\hat{\phi}}_c \nabla {\hat{\phi}}_c  - \nabla^{2} \hbar G_{cc} )   .
\end{eqnarray}

At this stage, we confine ourselves to the constant field 
configuration for the   $\hat{\phi} _{a}  (x)  $  fields, so  that  the  square   of the   
gradient of 
$\hat{\phi} _{a} (x) $ in Eq.(18) vanishes.
The scheme to expand $\langle \Psi | H | \Psi \rangle  $ 
in powers of $ \hbar  $ and discard terms higher than second order 
in $ \hbar^{2} $ will be adopted in the below.
The $ \hbar  $ expansion is equivalent to the loop expansion 
\cite{jus,nam}. Thus, we are going to study the system to two-loop order.
The energy density given in Eq.(18) can be expanded to second order 
in $ \hbar  $ to yield
\begin{eqnarray}
\langle   H   \rangle  &=& \frac{{   1}}{ 2}   \frac{{ g}}{    N} f^{2}  ( {\hat{\phi}} 
 ) 
\left[ 
1+\frac{{1}}{2}  \frac{{  \hbar   G_{aa}}}{f( \hat{\phi}    )} \right] \frac{{   \hbar 
}}{4   } 
G_{cc}^{-1} \nonumber  \\
            &  &-   \frac{{ 1}}{   2}   \frac{{   N}}{g  }f^{-2}  (   {\hat{\phi}} ) 
\left[1-\frac{{1}}{2} \frac{{ \hbar G_{aa}}}{f( {\hat{\phi}} )}  \right] 
 ( \nabla^{2} \hbar G_{cc} )   ,
\end{eqnarray}
where $f ( {\hat{\phi}} ) = 1+ {\hat{\phi}}^{2} /4$ has been defined.

We now vary this equation with respect to 
$G_{bb} (x,x) $ 
to determine the parameters $G_{bb} (x,x)$ that minimize the energy expectation value.
Thus, one gets the following relation:
\begin{eqnarray}
 \delta  \langle H\rangle &=& \frac{{   g}}{ N} \frac{{  \hbar    f( {\hat{\phi}} )}}{  
8} \left 
[\frac{{ 
\hbar  }}{2}  G_{aa}^{-1}- f( {\hat{\phi}} )[ 1 +\frac{{ \hbar }}{2 f( {\hat{\phi}} ) } 
G_{aa} ]^{2} \frac{ 1 }{ {G_{bb}}^{2} }  \right ] \delta G_{bb}  \nonumber \\
              & &+ \frac{{ N}}{g } \frac{{ \hbar }}{2 f^{2} ( {\hat{\phi}} ) } 
 \left [   - [ 1- \frac{{ \hbar  }}{2 f( {\hat{\phi}} ) } G_{aa} ] \nabla^{2}   +  
              \frac{{ \hbar }}{2 f( {\hat{\phi}} ) }   \nabla^{2} G_{aa} \right ]  \delta 
G_{bb}  \nonumber \\
               &=& 0 .
\end{eqnarray}
Here and in the   below, we will  use  the following notations.  The  repeated indices 
over  the 
letter $b $ does not indicate a sum; while, the repeated ones over $a $ means a sum.
This equation gives the relation that the variational parameter $G_{bb}$ must satisfy,
\begin{equation}
G_{bb}^{-2}  (x,y)  =  \left[-   \frac{{   N^{2}}}{g^{2} }    \frac{{4 }}{   f^{4}  (   
{\hat{\phi}}
)} \left( { 1- \frac{{ 3}}{2 }  \frac{{  \hbar  G_{aa}(z,z) }}{ f( {\hat{\phi}} )}}  \right)  
\nabla_{x}^{2} +  \frac{{ 1}}{2 }   \frac{{\hbar }}{ f( {\hat{\phi}} )  } 
G_{aa}^{-1}(z,z)  \right ] \delta (x-y),
\end{equation}
which has also been expanded  in  powers of $  \hbar  $ and 
terms higher  than $  \hbar  $ have  been discarded,   so  that only  terms up  to  $  
\hbar^{2}   $ 
can be retained in Eq.(19).
Since it is practically impossible to solve this equation directly, we 
separate the equation into two parts as follows: 
\begin{equation}
G_{bb}(x,y) = \frac{{ g}}{N } \frac{{  f^{2} ( {\hat{\phi}} ) }}{2  }
\frac{{ 1}}{[1-(3 \hbar  /2 f)  G_{aa} (z,z)  ]^{1/2}}   \int \frac{{  dp}}{2 \pi   } \frac{{  
1}}{ 
\sqrt{p^{2} +m^{2}} }{\exp }{[i p (x-y)]}
\end{equation}
and
\begin{equation}
m^{2} = \frac{{ g^{2}}}{N^{2} }  \frac{{ \hbar f^{3} ( {\hat{\phi}} )}}{8   }
  G_{aa}^{-1}(x,x) ,
\end{equation}
where the latter one turns out to define the mass parameter of the 
boson operator $\phi(x) $, as will be seen in Eq.(26).

We analyze the equation using the following iterative method.
 First, we approximate the   unknown  function   $G_{bb}(x,y) $    by 
$G_{bb}^{(0)} (x,y) $:
\begin{equation}
G_{bb}^{(0)} (x,y) = \frac{{ g}}{N } \frac{{  f^{2} ( {\hat{\phi}}  ) }}{2  } \int \frac{{ 
dp}}{2 \pi } \frac{{ 1}}{ \sqrt{p^{2} +m^{2}} }{\exp }{[i p (x-y)]}.
\end{equation}
Second, to improve this approximation, we substitute the equation back into the
coefficient of the right-hand side of Eq.(22).
Thus, we have
\begin{equation}
G_{bb}^{(1)} (x,y) = \frac{{ g}}{N } \frac{{  f^{2} ( {\hat{\phi}} ) }}{2  }
\frac{{ 1}}{[1-(3 \hbar /2 f) G_{aa}^{(0)} (x,x) ]^{1/2}}   \int \frac{{ dp}}{2 \pi  } \frac{{ 
1}}{ 
\sqrt{p^{2} +m^{2}} }{\exp }{[i p (x-y)]}.
\end{equation}
Note that the multiplicative factor in front of the integral has a divergence involving   a 
cutoff 
$\Lambda$. This must be removed by a proper renormalization of the wave function 
$\phi (x)$. 
Then, the  wave function renormalized expression for $ G_{bb}(x,y) $ becomes
\begin{equation}
G_{bb}^{(f)} (x,y) = \frac{{ g}}{N } \frac{{  f^{2} ( {\hat{\phi}}  ) }}{2  } \int \frac{{ 
dp}}{2 \pi } \frac{{ 1}}{ \sqrt{p^{2} +m^{2}} }{\exp }{[i p (x-y)]}.  
\end{equation}
This form will  be  used  in the  subsequent  discussions in  calculating the   energy 
expectation 
value. 
As a result of this renormalization, Eq. (21) is now rewritten
\begin{equation}
G_{bb}^{-2}  (x,y)  =  \left[-   \frac{{   N^{2}}}{g^{2} }    \frac{{4 }}{   f^{4}  (   
{\hat{\phi}}
)}  \nabla_{x}^{2}   +  \frac{{ 1}}{2 }   \frac{{\hbar }}{ f( {\hat{\phi}} )  } 
G_{aa}^{-1}(z,z)  \right ] \delta (x-y).
\end{equation}

Now, let us evaluate the Hamiltonian expectation value using Eq.(23) and (26).
The $ \nabla^{2} G_{aa} $ 
in the Hamiltonian can be calculated multiplying Eq.(27) by $ G_{bb}(y,z)$
 and integrating over the volume $\int dy $:
\begin{equation}
 \nabla^{2} G_{aa}(x,x)  = -  \frac{g^{2} }{N^{2}  } \frac{  f^{4}  ({ \hat{\phi}}
)}{4  } G_{aa}^{-1}(x,x)
 \left[ 1- \frac{1}{2} \frac{{ \hbar }}{  f({\hat{\phi}} )} G_{aa} (x,x) \right] .
\end{equation}
The hamiltonian density has no  classical  potential energy part $V(\phi)$,  so we  are 
allowed 
to set
\begin{equation}
f({\hat{\phi}} ) =1 .
\end{equation}
Thus, the Hamiltonian  to the second order   in $ \hbar   $, that is,  to  the two-loop  
order, 
is given by
\begin{eqnarray}
\langle \Psi | H | \Psi \rangle & = &                   
 \frac{{ g}}{   N}  \frac{{ \hbar     }}{8 } G_{aa}^{-1}(x,x) [1+   \frac{{ \hbar  }}{ 2 
} G_{aa} (x,x) ]   
            +  \frac{{  g}}{  N}    \frac{{ \hbar     }}{8   } 
G_{aa}^{-1}(x,x)[1-  \frac{{ \hbar  }}{ 2 } G_{aa} (x,x)  ] ^2 
\nonumber \\
        &=&  \frac{{   g}}{N  }   \frac{{  \hbar     }}{4  } 
G_{aa}^{-1}(x,x) [1-   \frac{{ \hbar  }}{ 4 } G_{aa} (x,x) ] 
 \end{eqnarray}

Using the  mass defining   relation Eq.(23)  and  two  point Green's function  Eq.(26),
one can evaluate $\langle  H  \rangle $,
\begin{eqnarray}
\frac{{\langle H \rangle  }}{N} & =& \frac{{  2m^{2}  }}{ g  }  (1- \frac{\hbar}{4}      
  G_{aa} (x,x)    )\nonumber \\
& =& \frac{ 2m^2  }{ g  } [1-  \frac
{ \hbar  g  }{  8}  \int_{- \Lambda}^{\Lambda}   \frac{   dp }{2  \pi   } \frac{{
1}}{   
\sqrt{p^{2} +m^{2}} }  ]  \nonumber \\
                &  =& \frac{{ 2m^{2}  }}{ g } [1-   \frac{{ \hbar g
}}{ 16 \pi} \ln ( \frac{{ 4 \Lambda^{2}}}{m^{2} } )  ],
\end{eqnarray}
where $ \Lambda  $ is an ultraviolet momentum cutoff.
To make the energy density finite, we may renormalize the parameter of the theory.
Defining the renormalized coupling constant $g_{r} $,
\begin{equation}
 \frac{{ 1}}{ g} =  \frac{{ 1}}{ g_{r}} +  \frac{{ \hbar  }}{ 16 \pi}  \ln 
 ( \frac{{ 4 \Lambda^{2}}}{\mu^{2}}   ) ,
\end{equation}
we obtain the finite energy density with the renormalization scale $\mu $:
\begin{equation}
\frac{{\langle H \rangle }}{N}  = \frac{{ 2m^{2} }}{ g_{r}  } + \frac{{ 
\hbar 
m^{2} 
}}{ 8 \pi} \ln( \frac{{ m^{2}}}{\mu^{2} } ).
\end{equation}
Note that  this is  the result up to $\hbar  ^2$, since  the mass  parameter $m^2$  is 
of 
$\hbar$ order. 
This energy expectation value has a minimum  away from the origin, since
it is concave upward as $m^{2} $ increases.  

Hence, by minimizing the energy expectation value with respect to 
$m^{2} $, one can derive the mass gap,
\begin{eqnarray}
 \frac{{  \partial   \langle H   \rangle /  N }}{\partial    m^{2} }  &  =   &   \frac{{ 
2}}{   
g_{r}  } 
+ \frac{{ \hbar }}{ 8 \pi} \ln ( \frac{{ e m^{2}}}{\mu^{2} } ) \nonumber \\
                                           &=& 0.
\end{eqnarray}
From this relation, one obtains the dynamically generated mass gap:
\begin{equation}
\langle m^{2} \rangle  = \mu^{2} \exp  \left [ -1-  \frac{{ 16 \pi}}{  \hbar g_{r}     }  
\right ] .
\end{equation}
The NLS model has no dimensional parameters; the coupling $g $ 
is dimensionless in two dimensions. But, we  arrived at  a  dimensional parameter 
$m^{2}   $; this phenomena  is an example of dimensional transmutation. 
At this value of the mass gap, the energy density to two-loop order becomes
\begin{eqnarray}
\frac{{\langle H   \rangle }}{N}  &   = &   \frac{{   2  \langle m^{2}   \rangle  }}{   
g_{r} } +  \frac{{ 
\hbar \langle   m^{2}   \rangle   }}{   8   \pi}   \ln (   \frac{{  \langle   m^{2}  
\rangle}}{\mu^{2}  }  ) \nonumber \\
                  &= &- \frac{{ \hbar  }}{8   \pi } \mu^{2} \exp \left  [-1 - \frac{{ 16
\pi}}{ \hbar g_{r}   }  \right ]  .
\end{eqnarray}
Note that the negative sign of the energy density indicates that the 
massive ground state is more stable than the massless one in the NLS model.

We now return to the mass defining equation, Eq.(23).
Differentiating it  with respect   to $m^{2}  $, one    finds the relation  between  the  
coupling 
constant $g$ and the cutoff $\Lambda$
\begin{eqnarray}
 \frac{{ 1}}{ g}  & =&   \frac{{ \hbar   }}{8 } \int \frac{{ dp}}{2 \pi 
} \frac{{ 1}}{ \sqrt{p^{2} +m^{2}} } \nonumber  \\
                  &= & \frac{{ \hbar  }}{ 16  \pi} \ln \left ( \frac{{ 4 
\Lambda^{2}}}{m^{2}}  \right ) .
\end{eqnarray} 
This relation shows  that as  $ \Lambda   $ becomes large,  the coupling  constant $g  
 $ 
approaches 
zero, thus satisfying the  asymptotic freedom. This   equation can be also  rewritten in 
terms of 
the renormalized coupling $g_{r}  $   and the renormalization  mass  scale $\mu $  in  
Eq.(32).

\section{Conclusion}
In this paper, the functional Schr\"{o}dinger picture approach has been applied to  
analyze the $O(N) $ nonlinear sigma (NLS) model.
We have considered the $O(N) $ NLS model in the 
large $N $-limit and calculated 
the energy expectation value to the second order in $ \hbar  $(two-loop order) 
systematically, using the 
gaussian wave functional form.
The Schr\"{o}dinger picture approach combined with the variational technique 
produced the   mass  gap  and the asymptotic freedom  of the ground state 
for the $O(N) $ NLS model in a straightforward manner.

The majority of the literature on the $O(N) $ NLS model adopt the 
Lagrangian formalism to investigate its nonperturbative phenomena, where a 
composite auxiliarly  field   $\sigma  (x) =\sum   _{a} \Phi_a  (x)   \Phi_a  (x)$ is    
usually 
introduced. 
However, here we discuss its nonperturbative phenomena directly without resorting to
the superfluous auxiliarly field.

The extension of our calculations to three dimensions will be straightforward.; 
the difference is  to write  the  Green's  function in  Eq.(26) in   corresponding three 
dimensional 
form and carry out the integral in the three dimension.

\vspace{0.5cm}
\noindent
{\large{\bf Acknowledgements}}
\vspace{0.5cm} \\

We thank Prof. K. -S. Soh and J. H. Yee for helpful discussions.
This  work   was supported   in   part   by the   Korea   Ministry   of Education 
(BSRL-96-2425),  the 
Korea Science   and Engineering   Foundation through  Project  95-0701-04-01-03  and 
through the 
SRC Program  of SNU-CTP.  Also,  D.  K. Kim  acknowledges the   Korea Research 
Foundation 
for support through the domestic postdoctorial program.


\begin{references}
\bibitem{nop}
J. Bardeen, L. N. Cooper, and J. R. Schrieffer, \prn 108, 1175 (1957);
Y. Nambu and G. Jona-Lasinio, \prn 122, 345 (1961).
\bibitem{vac}
R. Floreanini and R. Jackiw, \prd 37, 2206 (1988).
\bibitem{sig}
E. Br\'{e}zin and J. Zinn-Justin, \prd 14, 2615 (1976);
V.A. Novikov, M. A. Shifman, A. I. Vainshtein, and V. I. Zakharov, Phys. 
Rep. 116, 103 (1984); A. C. Davis, J. A. Gracey, A. J. Macfarlane, and M. 
G. Mitchard, \npb 314, 439 (1989).
\bibitem{bar}
W. A. Bardeen, B. W. Lee, and R. E. Shrock, \prd 14, 985 (1976);
T. del R. Gaztelurrutia and A. C. Davis, \npb 347, 319 (1990).
\bibitem{jus}
 J. Zinn-Justin, {\it Quantum Field Theory and Critical Phenomena } (Oxford 
University Press, New York, 1989). 
\bibitem{sch}
J. M. Cornwall, R. Jackiw, and E. Tomboulis, \prd 10, 2428 (1974);
R. Jackiw and A. Kerman, Phys. Lett. 71A, 158 (1979); 
A. Duncan, H. Meyer-Ortmanns, and R. Roskies, \prd 36, 3788 (1987).
\bibitem{res}
S. K. Kim, J. Yang, K. S. Soh, J. H. Yee, \prd 40, 2647 (1989); S. K. Kim, 
K. S. Soh, J. H. Yee, \ibd 41, 1345 (1990); G. Amelino-Camelia and S. -Y. Pi,
\ibd 50, 2356 (1994); G. Dunne, \ibd 50, 5321 (1994).
\bibitem{pis} 
S.-Y. Pi and M. Samiullah, \prd 36, 3128 (1987).
\bibitem{noh} 
H. S. Noh, C. K. Kim, and K. Nahm, Phys. Lett. 204A, 162 (1995); 
\ibd 210A, 317 (1996); J. Korean Phys. Soc. 29, 592 (1996); J. Song, S. Hyun, and  C. K. 
Kim, 
\ibd 29, 821  (1996); H. S.  Noh, S. K.  You, and C.  K. Kim, Int.  J. Mod. Phys.  B11, 
1829 
(1997).  
\bibitem{mem}
N. D. Mermin and H.  Wagner, \prl 17,  1133  (1966); S. Coleman, Comm.  Math. Phys.  
31, 259 
(1973).
\bibitem{hei}
F. D. M. Haldane, Phys. Lett. 93A, 464 (1983);
F. D. M. Haldane, \prl 50, 1153 (1983).
\bibitem{con}
P. A. M. Dirac, {\it Lectures on Quantum Mechanics } (Yeshiva University, 
New York,
 1964); M. Henneaux and C. Teitelboim, {\it Quantization of Gauge Systems } 
(Princeton University Press, New Jersey, 1992). 
\bibitem{col}
S. Coleman, {\it Aspects of Symmetry } (World Scientific, Cambridge 
University Press, Cambridge, 1985). 
\bibitem{nam}
Y. Nambu, \plb 26, 626 (1966).
\end{references}
\end{document}